# Mind Modeling: A ToM-Based Framework for Personalization

Cristina Gena[1,*]

[1]Dept. of Computer Science, University of Turin, Italy

**Abstract**

User modeling has traditionally relied on inferring preferences, traits, or intents from observable behaviour. While effective in many adaptive systems, this paradigm treats behaviour as the primary object of modeling and leaves mental-state attribution implicit. This assumption becomes limiting in socially situated and longitudinal interaction, where behaviour must be interpreted in context and over time. We introduce *mind modeling*, a perspective in which user modeling is grounded in the explicit and revisable attribution of mental states, including beliefs, intentions, emotions, and knowledge. Drawing on Theory of Mind (ToM), this approach treats behaviour as evidence for hypotheses about internal states, supporting personalization that is more interpretable and coherent across interaction episodes. We present M$^3$, a conceptual framework that integrates perception, mentalisation, and action within a unified structure, enabling the continuous update of mental-state hypotheses in embodied interaction. We further illustrate this perspective through an embodied interaction trace, providing an initial operationalization of mind modeling in practice.

## 1. Introduction

User modeling has traditionally been framed as user profiling, i.e., the inference of user preferences, traits, or intents from observable behaviour and interaction histories [1, 2, 3]. This paradigm has enabled effective personalization in many adaptive systems, including recommender systems, intelligent tutoring systems, conversational agents, as well as adaptive mobile and in-car services [4, 5]. In its dominant form, profiling assumes that correlations between observed behaviour and inferred user characteristics are sufficient to support meaningful adaptation.

However, this assumption becomes increasingly fragile in embodied, socially situated, and longitudinal interaction, such as those involving conversational agents [6] or social robots [7], where behaviour is context-dependent, socially grounded, ambiguous, and requires interpretation over time [8]. In these settings, users interpret system behaviour in terms of beliefs, intentions, and emotions [9, 10, 11], and personalization is evaluated not only by predictive accuracy, but also by coherence, interpretability, and social appropriateness over time. The same observable behaviour may correspond to different underlying mental states (e.g., hesitation as uncertainty, anxiety, or avoidance), making purely correlational models difficult to interpret and justify.

To address this limitation, we introduce *mind modeling*, a perspective in which user modeling is grounded in the explicit and revisable attribution of mental states. Mind modeling treats behaviour as evidence for hypotheses about beliefs, intentions, emotions, and knowledge, which are continuously updated across interactions. This perspective is particularly suited to embodied settings where perception and action are tightly coupled and does not replace profiling-based approaches, but complements them: behavioural regularities and preferences remain, yet they are interpreted within a broader inferential process aimed at explaining why a user behaves in a certain way.

*Joint Proceedings of the ACM UMAP Workshops 2026, UMAP 2026, June 8–11, 2026, Gothenburg, Sweden*
*Corresponding author.

Envelope-Open cristina.gena@unito.it (C. Gena)
GLOBE https://www.di.unito.it/~cgena/ (C. Gena)

Orcid 0000-0003-0049-6213 (C. Gena)

While the use of Theory of Mind (ToM) in artificial agents has been explored in AI and HCI [12, 13, 14], existing approaches are typically task-specific and do not frame ToM as a persistent, longitudinal user model for personalization. Our contribution is to position ToM as a foundational principle for user modeling itself, where mental-state attribution supports both adaptive behaviour and its explanation.

This paper makes three contributions: (i) it reconceptualizes user modeling as mental-state attribution grounded in interaction; (ii) it introduces M$^3$, a ToM-based conceptual framework integrating perception, mentalisation, and action; and (iii) it provides an initial operationalization through an embodied interaction trace and an implementation perspective, demonstrating how mental-state hypotheses can drive personalization and explanation over time.

## 2. Related Work

User modeling [1, 2], has long supported personalization in adaptive systems. Early approaches relied on symbolic and rule-based representations, while later work introduced probabilistic and data-driven methods, including Bayesian models [15], hidden Markov models [16], collaborative filtering [17], and deep learning techniques [18, 19]. These approaches have significantly improved scalability and predictive performance.

Despite their methodological diversity, most approaches share a correlational logic: they infer preferences, traits, or goals from observed behaviour without explicitly modeling the underlying mental states that generate such behaviour. As a result, adaptation is often driven by patterns in the data rather than by an explicit account of beliefs, intentions, or emotions. This limitation becomes more evident in socially situated and longitudinal interaction, where behaviour is context-dependent, ambiguous, and requires interpretation over time.

These challenges are particularly visible in embodied AI and social robotics, where systems operate in shared physical and social environments and interact through multiple modalities. In this context, user modeling must account not only for preferences, but also for situated action, affect, and interaction dynamics. Existing frameworks, such as the three-layer model by Rossi et al. [20], distinguish physical, cognitive, and social dimensions of user profiling, supporting tasks such as intention recognition, personality inference, and emotion detection. However, these approaches remain largely reactive and task-specific, and do not provide a persistent, revisable representation of the user as an intentional agent.

Theory of Mind (ToM), defined as the capacity to attribute mental states to oneself and others [21], provides a principled foundation for addressing this gap. ToM has been extensively studied across disciplines [22, 23], and includes both cognitive and affective components, supporting reasoning about beliefs, intentions, and emotions. There is no consensus on how humans implement Theory of Mind. Theory-Theory posits that individuals rely on an implicit folk-psychological theory to infer others' mental states [24, 21]. Simulation Theory instead suggests that people understand others by internally simulating their possible mental processes [25, 26]. In contrast, direct-perception and interaction-based approaches argue that some mental states can be directly perceived in embodied interaction, without full inferential reconstruction [27, 28]. Rather than committing to a single account, the framework proposed here is intentionally compatible with these perspectives. Theory-like components motivate explicit mental-state representation; simulation motivates the role of self-related representational resources in modeling others; and direct-perception highlights the importance of embodied and interactional cues, cautioning against treating all mental-state attribution as purely inferential. In this sense, M$^3$ can be interpreted as a unifying framework in which mental-state attribution is grounded in perception, supported by internal representational resources, and continuously refined through interaction.

Recent work in AI and HCI has explored ToM-inspired mechanisms in artificial agents [12, 13, 14, 29]. However, these approaches are typically task-specific or focused on agent interaction, and do not frame ToM as the basis of a persistent user model for personalization. In contrast, this paper proposes to treat user modeling itself as a process of explicit and revisable mental-state attribution, where behaviour is interpreted as evidence and personalization is grounded in the evolving representation of the user's mind.

## 3. From User Modeling to Mind Modeling

In this paper, we define *mind modeling* as a form of ToM-based user modeling that maintains explicit, revisable hypotheses about user mental states, including beliefs, goals, intentions, emotions, and knowledge. These hypotheses are continuously updated on the basis of interaction evidence and are used to guide both personalization decisions and their explanation. Rather than treating behaviour as the primary object of modeling, mind modeling treats behaviour as the observable manifestation of underlying mental states that evolve across interaction episodes. This shift entails three consequences.
*Representational consequence.* Mind modeling represents users through structured models of mental states that are dynamic, uncertain, and context-dependent. User preferences and behavioural regularities are not discarded, but repositioned as evidence or secondary descriptors rather than the core ontology of the user model.
*Inferential consequence.* Personalization moves from direct mapping over behavioural traces to interpretative reasoning, where the same behaviour may support multiple competing hypotheses. Instead of assuming that one pattern implies one stable user characteristic, the system maintains alternative mental-state explanations and revises them as evidence accumulates.
*Action consequence.* Personalization becomes not only reactive but anticipatory and coherent over time. The system can justify not only what it did, but why it did it, by referencing the hypotheses that grounded the decision and their associated uncertainty.

This is precisely where mind modeling differs from both standard user modeling and conventional belief tracking. Probabilistic user models often include latent variables, but these are usually taskdependent abstractions introduced to improve behavioural prediction. In contrast, mind modeling treats beliefs, goals, and emotions as ontologically explicit mental-state hypotheses introduced to explain behaviour, not merely predict it. Moreover, belief tracking is often episodic and local, whereas mind modeling is intrinsically longitudinal, supporting revision and continuity across interaction episodes.

### 3.1. M$^3$: A ToM-based Framework

M$^3$ is a theory-based framework that specifies the representational and inferential commitments required for mind modeling. M$^3$ can be naturally interpreted within an embodied perception-action loop, where mental state attribution is grounded in interaction. In this perspective, mind modeling is not a static inference process but a continuous cycle in which the system perceives the user, updates hypotheses about the user's mental states, and produces situated actions that, in turn, shape future interaction. M$^3$ thus operates as the cognitive layer of an embodied agent, coupling perception, mentalisation, and action over time.

The framework introduces two coupled representational spaces: (i) an internal space encoding the system's own interaction-relevant states, and (ii) a user-directed space representing hypothesized user mental states with associated uncertainty. This distinction reflects the metacognitive nature of ToM,
where attributing mental states to others presupposes internal representational resources.

In embodied settings, interaction evidence is grounded in multimodal perception, see for details [30, 31]. Let $o_t$ denote raw observations at time $t$, including language, visual cues (e.g., gaze, facial expression), and behavioral signals (e.g., hesitation, actions). These observations are mapped to structured evidence:

$$E_t = \text{Percept}(o_t)$$

which serves as input to the mind modeling process.

The user mind model is then updated through a mentalisation operator:

$$m_{t+1} = U(m_t, E_t, s_t)$$

where $m_t$ represents current hypotheses about user mental states, $E_t$ is perceptual evidence, and $s_t$ denotes the system's internal state.

Personalization is realized through situated action:

$$a_t = \pi(m_t, s_t, c_t)$$

where $a_t$ corresponds to embodied behaviors such as speech or gesture and $c_t$ encodes contextual constraints.

This process defines a closed loop: system actions affect the environment and the user, generating new observations $o_{t+1}$, which in turn produce new evidence $E_{t+1}$. As a result, mental state hypotheses are continuously revised in response to interaction dynamics.

M$^3$ can be described through three core elements.

**Mental state space.** Let $M$ be a structured space of mental state variables (beliefs, goals, intentions, emotions, knowledge). The user model at time $t$ is a state $m_t \in M$ with associated uncertainty. **Perception and evidence.** Raw observations $o_t$ are mapped to interaction evidence $E_t$, capturing signals relevant to mental state attribution.

**Update and action.** The system revises its hypotheses through an update operator $U$, and selects actions through a policy $\pi$, grounding decisions in explicit mental state representations.

This formulation highlights that, in M$^3$, behavior is not predicted directly from data but interpreted through hypothesized mental states that evolve over time. Embodiment ensures that these hypotheses remain anchored to perceptual signals and interaction context, enabling adaptive and explainable behavior in situated settings.

To illustrate how M$^3$ operates in practice, consider an assistive robot interacting with an elderly user during a daily medication routine.

$t_0$ **- Initial state.** The system maintains uncertain hypotheses about the user's beliefs regarding medication intake (e.g., whether the user believes the medication has already been taken). **Perception.** The robot observes that the user does not mention the medication during a routine check-in and exhibits slight hesitation:

$$o_t \rightarrow E_t = \{\text{omission, hesitation}\}$$

**Mentalisation update.** Rather than inferring a generic state (e.g., low adherence), the system updates competing hypotheses:

- $h_1$: the user believes the medication was already taken
- $h_2$: the user is uncertain or has forgotten
- $h_3$: the user is avoiding the topic

Each hypothesis is associated with a confidence value and revised as evidence accumulates. **Action.** The system selects a context-sensitive action:

$$a_t = \pi(m_t, s_t, c_t)$$

For instance, it may ask: *“I might be mistaken, but I was wondering if you already took your medication today?”*

**Loop closure.** The user’s response generates new observations $o_{t+1}$, leading to further updates of the mental state model. This trace highlights the key difference from profiling-based approaches: behavior is not treated as direct evidence of a fixed trait, but as ambiguous data requiring interpretation. Embodied interaction provides the signals that ground this interpretation, while $M^3$ structures how these signals are transformed into revisable mental state hypotheses and actionable decisions. This abstraction provides the basis for concrete architectural instantiations, as outlined next.

### 3.2. Implementation Perspective

Building on these theoretical premises, we are currently implementing the approach within a hybrid cognitive architecture inspired by the Common Model of Cognition [32], adopting memory and control mechanisms consistent with architectures such as SOAR, and extending the architecture with ToM as a new pillar for socially credible and personalized interaction. The proposed cognitive architecture integrates symbolic reasoning, subsymbolic perception, and embodied simulation, supported by SOARinspired memory subsystems. Dual knowledge sources, namely large language models with RetrievalAugmented Generation (RAG) and declarative ontologies, jointly ensure scalability, contextualisation, and transparency of reasoning. At its core, a User Modeling Reasoner operationalises the Attribution of Mental States, constructing probabilistic user models from multimodal signals. This architecture instantiates the two representational spaces introduced in M3: an internal state space supporting reasoning and interaction management, and a user-directed space maintaining probabilistic hypotheses about the user’s mental states. *Memory subsystems.*

- **Working Memory** maintains the current interaction state and live ToM predicates (e.g., *userbelieves(p)*, *user-feels(e, intensity)*).
- **ProceduralMemory** stores policies governing sensorimotor routines and higher-level interaction strategies for prosocial behaviour.
- **Semantic (Declarative) Memory** encodes ontologies of mental states and domain knowledge, ensuring explicit and transparent representations [33].
- **Episodic Memory (Bayesian)** encodes interaction experiences and supports belief revision through Bayesian updating [34].

*Dual knowledge sources.*

1. **LLMs with RAG**, supporting dialogue and explanation through contextualised knowledge retrieval.
2. **Declarative ontologies**, enabling traceable and interpretable reasoning.

*Cognitive layers.*

- **Symbolic Reasoning Layer**: performs ToM reasoning, belief revision, and perspective-taking.
- **Subsymbolic Perception Layer**: extracts features from multimodal input using deep learning techniques.
- **Embodied Simulation Layer**: supports synchrony, mirroring, and affective resonance.

*Bayesian episodic loop.* A central mechanism integrates perception, reasoning, and memory through continuous updating of probabilistic user models, enabling both short-term adaptation and long-term personalization.

*User modeling module.* Within this architecture, a central component is the User Modeling module, which operationalises mind modeling by maintaining and updating mental-state hypotheses. It is inspired by the AMS framework [35], representing epistemic, emotive, intentional, imaginative, and perceptual dimensions with confidence scores in [0, 1], structured as follows:

- **Input**: multimodal cues (e.g., facial expressions, gaze, prosody, posture, language).
- **Representation**: mental-state dimensions modeled as latent variables.
- **Confidence**: probabilistic estimates reflecting uncertainty.
- **Update**: belief revision through episodic memory.
- **Use**: adaptive behaviour and explanation generation.

In this setting, user preferences and behavioural regularities are not represented as primary model variables, but are treated as evidence derived from interaction history, contributing to the revision of mental-state hypotheses. This evidence is processed by the User Modeling Reasoner module, which maintains and updates a mental-state representation $s$ with associated uncertainty. This representation is then used to guide behaviour selection and language generation, ensuring alignment between inferred user states and system outputs.

## 4. Conclusion and Future Work

Mind modeling reframes explainability as a structural property of personalization rather than an addon. Because system behaviour is grounded in explicit mental-state hypotheses and their associated uncertainty, decisions can be traced and justified over time. This also supports longitudinal coherence, allowing the system to explain not only individual actions but also changes in behaviour across interaction episodes.

This perspective has implications for evaluation. While profiling-based systems are typically assessed through predictive accuracy, mind modeling requires additional dimensions, including: (i) coherence across episodes, (ii) plausibility of mental-state revision, and (iii) explanation helpfulness [36]. These dimensions can be operationalized through longitudinal analysis of state updates, agreement with human judgments or domain experts, and established HCI measures of clarity, trust, and perceived justification.

Mind modeling enables more adaptive and transparent interaction, but also introduces risks, including over-interpretation, privacy intrusion, and unwarranted confidence in inferred mental states. Systems may attribute beliefs or emotions that users have not explicitly expressed, or act on uncertain hypotheses in ways that influence user behaviour. These risks require explicit uncertainty representation, conservative update strategies, and transparency about what is inferred and why, together with mechanisms for users to understand and contest system assumptions.

This work represents an initial step toward operationalizing mind modeling and has several limitations. First, $M^3$ does not prescribe a unique computational implementation, and different realizations of $U$ and $\pi$ are possible. Second, ToM constructs are complex and their operationalization varies across domains and measures [21, 37]. Finally, while an initial implementation is underway, this paper does not provide empirical validation, but aims to establish a conceptual and architectural foundation for future work.

Ongoing work focuses on implementing and evaluating this approach in real interaction scenarios, assessing its impact on adaptation, explainability, and user trust in socially situated settings such as assistive robotics.

## Declaration on Generative AI

The author employed ChatGPT exclusively for orthographic and grammatical proofreading of the final manuscript draft. After using this tool, the author reviewed and edited the content as needed and takes full responsibility for the publication’s content.